\documentclass[multphys,vecphys]{svmult}

\usepackage{makeidx}     % allows index generation
\usepackage{graphicx}    % standard LaTeX graphics tool
\usepackage{multicol}    % used for the two-column index

\makeindex             % used for the subject index
\newcommand{\snia}{SN~Ia}
\newcommand{\sneia}{SNe~Ia}
\newcommand{\ra}{$\rightarrow$}
\begin{document}

\title*{Synthetic Spectra for Type~Ia Supernovae\\ at
  Early Epochs}
\titlerunning{Synthetic Spectra for Type~Ia Supernova}

\author{D.N.~Sauer\inst{1}\and A.W.A.~Pauldrach\inst{2} \and
  T.~Hoffmann\inst{2} \and W.~Hillebrandt\inst{1}}
\institute{Max-Planck-Institut f\"{u}r Astrophysik,
  Germany\\\texttt{dsauer@mpa-garching.mpg.de,
  wfh@mpa-garching.mpg.de} 
\and Institut f\"{u}r
  Astronomie und Astrophysik der Universit\"{a}t M\"{u}nchen, Germany\\
  \texttt{uh10107@usm.uni-muenchen.de, hoffmann@usm.uni-muenchen.de}}
%
% Use the package \"{u}rl.sty" to avoid
\maketitle

\index{NLTE model}
\index{SNIa}
\index{synthetic spectrum}
\index{radiative transfer}

\abstract

We present the current status of our construction of synthetic spectra for
type Ia supernovae.  These properly take into account the effects of NLTE
and an adequate representation of line blocking and blanketing.  The
models are based on a sophisticated atomic database.  We show that the
synthetic spectrum reproduces the observed spectrum of 'normal'  SN-Ia
near maximum light from the UV to the near-IR.  However, further
improvements are necessary before truly quantitative analyses of observed
SN-Ia spectra can be performed.  In particular, the inner boundary
condition has to be fundamentally modified. This is due to the dominance
of electron scattering over true absorption processes coupled with the
flat density structure in these objects.

\section{Introduction}

A detailed understanding of the physics and the explosion mechanism of
type~Ia supernovae (\sneia) is essential to minimizing systematic
errors of cosmological parameters derived from the interpretation of
{\sneia} observations. A key role in this context play realistic
models of the expanding atmospheres of {\sneia} which are based on the
full non-equilibrium physics involved, which link the results of
current hydrodynamic explosion models (see, e.g., Reinecke et al.~2002
\cite{reinecke02}, Gamezo et al.~2003 \cite{gamezo03}) to the
comprehensive observational data.

The spectra of {\sneia} contain useful information about the
energetics of the explosion---luminosity and velocity---and the
nucleosynthesis, especially in shells above the pseudo-photosphere,
which are primarily observable in early epochs. To gain access to this
information several groups have started to compute synthetic optical
spectra for {\sneia} in the photospheric epoch (Pauldrach et al.~1996
\cite{pauldrach96}, Nugent et al.~1997 \cite{nugent97}, Lentz et
al.~2001 \cite{lentz01}, H\"oflich et al.~2002 \cite{hoeflich02}); so
far, however, all use simplifying assumptions with uncertain influence
on the resulting spectrum.  Realistic models for {\em quantitative}
analysis are still missing.

The main reason we still do not have realistic models is that the
physical conditions in expanding {\snia}-atmospheres---high radiation
energy density in a low matter density environment---make this
modeling especially difficult. Simplifying assumptions such as local
thermodynamic equilibrium (LTE) are invalid because radiative
processes dominate over local collisional processes.  Thus, a
realistic model requires at minimum a solution of the full non-LTE
problem.  Other difficulties arise because the ejecta do not contain
significant amounts of hydrogen or helium, elements that form the main
contribution to continuous opacity in other stellar objects. This
causes a strong dominance of line opacity over continuous opacity,
which is further complicated by the high expansion velocities of the
ejecta (up to $\sim 30000\,$km/s). Thus, thousands of Doppler-shifted
and -broadened spectral lines of low-ionized metal atoms overlap and
form a non-thermal ``pseudo-continuum'' with strong line-blocking and
line-blanketing effects.

Furthermore, the entire light emission is powered by $\gamma$-photons
(and in later times also positrons) originating from the decay of
$\,^{56}$Ni~{\ra}~$^{56}$Co~{\ra}~$^{56}$Fe. These $\gamma$-photons
are redistributed to lower energies by scattering and ionization
events.  They are unlikely to be thermalized---in best case only
partially---even within the pseudo-photosphere and the outer-most
parts of the atmosphere because the density increases only moderately
toward the center.  Recent explosion models predict that $^{56}$Ni can
be present even at high velocities (Reinecke et al.~2002
\cite{reinecke02}). This indicates the need for a consistent treatment
of the energy input within the atmosphere, which may significantly
influence the emergent spectrum.  Also, the observed changes in the
spectra with time as the photosphere recedes into the ejecta show that
the abundance of the elements varies strongly with radius.

All these factors have to be taken into consideration to develop a
realistic radiative transfer model that allows a reliable quantitative
analysis of {\sneia}.

In this paper, we describe the computation of synthetic spectra for
{\sneia} based on a consistent treatment of the full NLTE radiative
transfer (see Sauer \& Pauldrach 2002 \cite{sauer02}).  This method
will be used to test hydrodynamic explosion models against the
observations and will eventually provide a powerful tool for spectral
diagnostics of {\sneia}.

\section{The Method}
The basic approach for modeling supernova spectra is analogous to the
treatment of O-stars with expanding atmospheres because the physics
involved is similar in many respects. Our code development is based on
a program package which has proven to be very successful for
the quantitative spectral analysis of hot stars with radiation-driven
winds (Pauldrach et al.~2001 \cite{pauldrach01}, Pauldrach 2003
\cite{pauldrach03}).

The models treat the expanding atmosphere as stationary and
spherically symmetric.   The underlying
density structure, velocity field, and composition of the ejecta
(averaged over angles) are provided by the explosion model.

The code provides a consistent solution of the radiative transfer and
the NLTE rate equations for the occupation numbers of the atomic
levels. A proper treatment of line blocking and blanketing effects is
included.  All significant contributions to opacity and emissivity
from bound-bound, bound-free, and free-free transitions, as well as
Thomson-scattering, are taken into account---for this purpose we have
compiled and computed a comprehensive atomic database involving
accurate radiative and collisional ionization and excitation
cross-sections and atomic level energies of all important ions.

After each iteration cycle, the temperature for each radius point is
determined by balancing all energy gains and losses to the
gas---implementation of the $\gamma$-photon deposition is currently
under development; thus far, it is only considered with the assumption
that all deposition takes place below the photosphere. Together with
the use of the diffusion approximation at the the lower boundary this
implies the assumption of a complete thermalization of the
$\gamma$-photons.  The total luminosity (given by the amount of
synthesized $^{56}$Ni and the $\gamma$-deposition efficiency) is a
parameter that is used to fit the shape of an observed spectrum.

Finally, the emergent spectrum is given by the computed radiation
field at the outer boundary of the converged model.

\section{First Results}

\begin{figure}
  \centering
  \includegraphics[angle=-90,width=.7\textwidth]{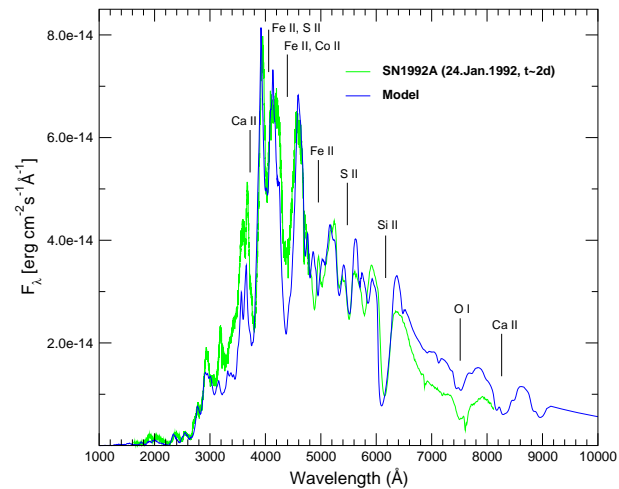}
  \caption{Synthetic spectrum for a {\snia} around maximum light based
    on the W7 explosion model with homogenized composition. For
    comparison, the observed spectrum of SN1992A is also plotted.}
  \label{dns:spectrum}
\end{figure}

Fig.~\ref{dns:spectrum} (see Sauer \& Pauldrach 2002 \cite{sauer02})
compares the synthetic spectrum of a NLTE-model based on the W7
explosion model (Nomoto et al.~1984 \cite{nomoto84}) to the observed
spectrum of SN~1992A.  At first glance the spectra seem to agree quite
well. However, a closer look reveals a few discrepancies: the model
reproduces the UV and blue part of the observed spectrum fairly well,
but it fails to match the relatively steep decline toward infrared
bands.  Additionally, the absorption features of the model that are
deeper than the observed may suggest that the peak fluxes are too
large.

Thus, we cannot yet consider the models to be sufficient for
quantitative spectroscopy of {\sneia}. Relevant {\snia} physics, in
particular the treatment of the $\gamma$-deposition and a better
analytical approach to the boundary condition at the photosphere (see
next section) still have to be included into the model.

\section{Discussion and Conclusions}

The enhanced fluxes of the model redward of $\sim 4000\,$\AA\ shown in
Fig.~\ref{dns:spectrum} result from the invalid assumption of a
photosphere analogous to the atmospheres of hot stars.
\begin{figure}
  \centering
  \includegraphics[width=.7\textwidth]{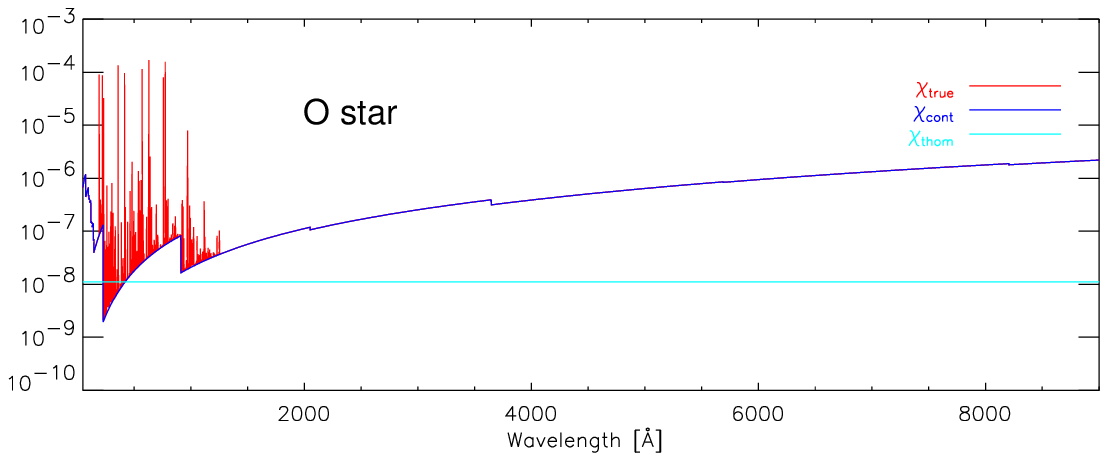}
  \includegraphics[width=.7\textwidth]{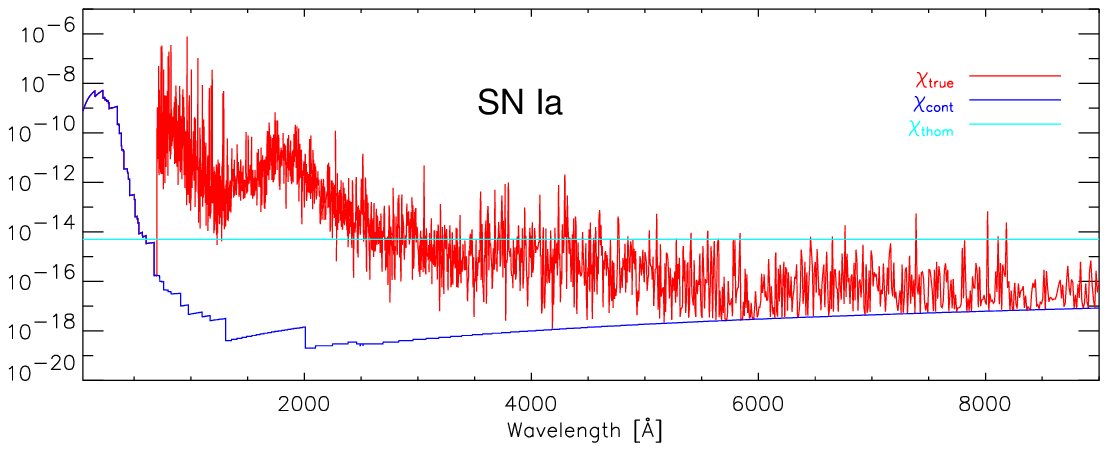}
  \caption{The contibutions to the opacity in a stellar atmosphere
    (upper panel) and a supernova (lower panel). Shown are the in both
    cases the Thomson-opacity $\chi_{\rm Thom}$ (the grey straight
    line) and the true continuum opacity $\chi_{\rm cont}$ (the dark
    solid line). $\chi_{\rm true}$ denotes the sum of the continuum
    opacity and the line opacity. }
  \label{dns:opa}
\end{figure}

Various conditions in the ejecta contribute to the breakdown of the
diffusion-approximation at the lower boundary. For stellar
atmospheres, the diffusion limit for radiation is an excellent
approximation as the conditions deep in the photosphere are very close
to LTE. This is because the opacities at all wavelengths are dominated
by true processes: H and He provide a strong bound-free continuum and
free-free processes are supported by the exponential density increase
at the photosphere (see Fig.~\ref{dns:opa}).

In {\sneia}, however, the opacity is dominated by electron scattering
over a large wavelength range, between the optical red and the radio
band (see Fig.~\ref{dns:opa})---the sources of true continuous opacity
(free-free and bound-free) are orders of magnitude smaller because the
composition does not contain hydrogen and helium, and the density
distribution is flat.  The electron scattering opacity, in contrast to
true opacities, does not couple the radiation field to the local
temperature.  As the flux at the inner boundary from the diffusion
approximation is determined by the temperature gradient which is
constrained only by the true processes and does not consider
scattering dominance.  The input flux at these wavelengths is
overestimated.  We note these conditions do not change at larger
optical depths.  Moreover, the situation already occurs at early
epochs where the assumption of a photosphere is generally considered
to be a valid approximation.

These results show that the commonly used lower boundary condition
(i.e., analytical expressions based on the diffusion-approximation) is
highly problematic for {\sneia}.  An inner boundary that correctly
accounts for the physical conditions is required in order to correctly
solve for the radiation field in {\sneia}.

Overall, the first results obtained with our method are encouraging.
The main spectral features of {\snia} spectra are well reproduced,
indicating that the underlying physics is correctly described.
However, the results also show that even at early epochs the {\sneia}
envelopes are comparable neither to the atmospheres of hot stars nor
to gaseous nebula. Thus, the assumptions usually made for these cases
cannot be applied to the expanding envelopes of {\sneia}. In
particular, a valid analytic approach to the boundary condition at the
inner radius of a scattering-dominated atmosphere has to be
formulated.

\subsubsection{Acknowledgements}
We thank our collaborators in the MPA Hydro-Group and
the USM Hot-Star group for helpful discussions. This work was
supported by the ``Sonderforschungsbereich 375-95 f\"ur
Astro-Teilchenphysik'' der Deutschen Forschungsgemeinschaft.
Attending the meeting was made possible by the EU CORDIS program.

%%%%%%%%%%%%%%%%%%%%%%%% referenc.tex %%%%%%%%%%%%%%%%%%%%%%%%%%%%%%
% sample references
% "physics"
%
% Use this file as a template for your own input.
%
%%%%%%%%%%%%%%%%%%%%%%%% Springer-Verlag %%%%%%%%%%%%%%%%%%%%%%%%%%

%
% BibTeX users please use
% \bibliographystyle{}
% \bibliography{}

\begin{thebibliography}{99.}
% Use the following syntax and markup for your references
%
% Monographs
%% \bibitem{monograph} H. Ibach, H. L\"uth: \textit{Solid-State
%% Physics}, 2nd edn (Springer, Berlin Heidelberg New York 1996) pp 45--56

%% % Contributed Works
%% \bibitem{contribution} D.M. MacKay: Visual stability and voluntary eye
%% movements. In: \textit{Handbook of Sensory Physiology}, vol 3, ed by R.
%% Jung, D.M. MacKay (Springer, Berlin Heidelberg New York 1973) pp
%% 307--331

\bibitem{gamezo03}
 V.N.~Gamezo, A.M.~Khokhlov, E.S.~Oran, A.Y.~Chtchelkanova, and
 R.O.~Rosenberg: Science, {\bf 299}, 77 (2003)

\bibitem{hoeflich02}
  P.~{H{\" o}flich}, C.L.~{Gerardy}, R.A.~{Fesen}, and  S.~{Sakai}:
  ApJ {\bf 568}, 791 (2002)

\bibitem{lentz01}
  E.J.~Lentz, E.~Baron, D.~Branch, and P.~Hauschildt: ApJ, {\bf 557},
  266 (2001)

\bibitem{nomoto84}
K.~{Nomoto}, F.~{Thielemann}, and K.~{Yokoi}: ApJ, {\bf 286}, 644 (1984)

\bibitem{nugent97}
P.~Nugent, E.~Baron, D.~Branch, A.~Fisher, and P.H.~Hauschildt: ApJ,
{\bf 485}, 812 (1997)

\bibitem{pauldrach03} A.W.A.~Pauldrach: Rev. Mod. Astron. {\bf 16}, in
  press (astro-ph/0202226) (2003)

\bibitem{pauldrach01}
 A.W.A.~Pauldrach, T.L.~Hoffmann, and M.~Lennon: A\&A, {\bf 375},
 161 (2001)

\bibitem{pauldrach96}
  A.W.A.~Pauldrach, M.~Duschinger, P.A.~Mazzali, J.~Puls, M.~Lennon,
  and D.~Miller: A\&A, {\bf 312}, 525 (1996)

\bibitem{reinecke02}
M.~Reinecke, W.~Hillebrandt, and J.C. Niemeyer: A\&A, {\bf 391}, 1167 (2002)

\bibitem{sauer02}
  D.~Sauer and A.W.A.~Pauldrach: Model atmospheres for type Ia
  supernovae: basic steps towards realistic synthetic spectra. In:
  Nuclear Astrophysics, ed. by W.~Hillebrandt and E.~M\"uller (MPA
  Proceedings, 2002) p 48
%% % Journal
%% \bibitem{journal} S. Preuss, A. Demchuk Jr, M. Stuke et al: Appl. Phys.
%% A \textbf{61}, 33 (1995)

%% % Theses
%% \bibitem{thesis} D.W.  Ross: Lysosomes and storage diseases. MA
%% Thesis, Columbia University, New York (1977)

\end{thebibliography}

%%% Local Variables: 
%%% mode: latex
%%% TeX-master: "dsauer_main"
%%% TeX-master: "dsauer_main"
%%% TeX-master: "dsauer_main"
%%% End: 

%%%%%%%%%%%%%%%%%%%%%%%%%%%%%%%%%%%%%%%%%%%%%%%%%%%%%%%%%%%%%%%%%%%%%%  }

%%%%%%%%%%%%%%%%%%%%%%%%%%%%%%%%%%%%%%%%%%%%%%%%%%%%%%%%%%%%%%%%%%%%%%

\printindex
\end{document}